# Solute-solute interactions in intermetallic compounds

Debashis Banerjee[*], Ryan Murray, and Gary S. Collins

*Department of Physics and Astronomy, Washington State University, Pullman, WA 99164, USA*

## Abstract

Two types of solute-solute interactions are investigated in this work. Quadrupole interactions caused by nearby Ag-solute atoms were measured at nuclei of [111]In/Cd solute probe atoms in the binary compound $GdAl_2$ using the method of perturbed angular correlation of gamma rays (PAC). Locations of In-probes and Ag-solutes on both Gd- and Al-sublattices were identified by comparing site fractions in Gd-poor and Gd-rich $GdAl_2(Ag)$ samples. Interaction enthalpies between solute-atom pairs were determined from temperature dependences of observed site fractions. Repulsive interactions were observed for close-neighbor complexes In/Gd/+Ag/Gd/ and In/Gd/+Ag/Al/ pairs, whereas a slightly attractive interaction was observed for In/Al/+Ag/Al/. Interaction enthalpies were all in the range ±0.15 eV. Temperature dependences of site fractions of In-probes on locally defect-free Gd- and Al-sites yields a transfer enthalpy that was found to be 0.343 eV in a previous study of undoped $GdAl_2$. The corresponding values in $GdAl_2(Ag)$ samples are much smaller. This is attributed to competition of In- and Ag-solutes to occupy sites of the same sublattice. While the difference in site-enthalpies of In-solutes on Gd- and Al-sites is temperature independent, it is proposed that the transfer of Ag-solutes from Gd- to Al-sites leads to a large temperature dependence of degeneracies of levels available to In-solutes, resulting in an effective transfer enthalpy that is much smaller than the difference in site-enthalpies.

---

[*] Permanent address: Accelerator Chemistry Section, RCD(BARC), Variable Energy Cyclotron Centre, Kolkata 700064, India

# Introduction

There is considerable interest in the site preference of solute atoms in compounds. A previous study was carried out of indium solute atoms in $GdAl_2$ using perturbed angular correlations of gamma rays, or PAC, as a function of the detailed composition and temperature of the compound [1]. $GdAl_2$ has the cubic $MgCu_2$ Laves structure and is highly ordered. The Gd-sites have cubic point symmetry and the Al-sites have axial symmetry. This allows one to readily distinguish occupations of the two sites by measuring nuclear quadrupole interactions. Measurements revealed that the site fraction of In probe atoms on Gd-sites, $In_{Gd}$, increases in Gd-poorer samples, consistent with the heuristic rule that impurity atoms tend to occupy the sublattice of an element in which there is a deficiency [1]. In addition, In-probes in thermal equilibrium were observed to transfer from Gd- to Al-sites with increasing temperature, in effect forming a two-level quantum system in which the enthalpy of a probe on the Al-site was apparently 0.343(3) eV higher than on the Gd-site [1].

A second subject of interest in the theory of solutions is the study of interactions between pairs of solute atoms. Interactions between solute atoms in pure metals were studied extensively in the 1980's by Krystof Królas and coworkers using PAC [2, 3, 4, 5 ] and by others using Mössbauer effect [6, 7]. Metals have a unique lattice location and interactions were observed, for example in Ag-metal, between a PAC probe such as $^{111}$In/Cd and close neighbor solutes such as Pd, In, or Sn. The $^{111}$In PAC probe was highly dilute (mole fraction $\sim 10^{-11}$). A favorable solute mole fraction used was $\sim$1 at.%, so that in the absence of interaction between the probe and solute, there would be a $\sim$10% site fraction for probes having a single solute neighbor, which can be readily measured. If the probe-solute interaction is attractive, the probe-solute site fraction will be large at low temperature and decrease with increasing temperature; if it is repulsive, the fraction will be small at low temperature and increase with temperature. Interaction enthalpies were determined from temperature dependences of the ratio of the site fractions of probes having one or zero neighboring solutes:

$$\frac{f_1}{f_0} \cong zc\exp(S/k_B)\exp(-Q/k_BT), \qquad\qquad (1)$$

in which z is the number of near neighbors, c is the solute mole fraction, and S and Q are the vibrational entropy and enthalpy of interaction. (*Q* is *negative* in the case of an attractive interaction.) Królas was able to explain observed interaction enthalpies as well as electric-field gradients caused by the neighboring solute atoms using a model of screened coulomb potentials [3]. Effective charges of solutes based on the difference between the nominal valence of a solute and the host atom it replaced proved useful to explain the results. Consider, for example, the interaction between In probes and Pt solutes in Ag [2]. Using nominal valences of +3, 0, and +1 for In, Pt and Ag, Królas obtained effective charges of +2 for In and -1 for Pt relative to the Ag-host atoms replaced, resulting in an *attractive* screened interaction. Likewise, probe and solute atoms exhibited repulsive interactions when both had higher valences than host atoms.

In the present work, the same approach is applied to investigate solute interactions in ordered binary compounds. Unlike in pure metals, an impurity atom will exhibit a preference to occupy sites of one element or the other. The effective charge of an impurity atom will now depend on the sublattice it occupies. This applies to both solute and probe atoms, so that the interaction enthalpy will differ depending on the sites occupied. This is illustrated in the present work, in which the interaction between In and Ag impurities in $GdAl_2$ is variously attractive or repulsive. $GdAl_2$(Ag) was chosen for study because the site preference of [111]In probes had previously been extensively studied and because it was found that In has an appreciable occupation of both sublattices [1]. The systems studied are strictly quaternary compounds, but the mole fraction of [111]In/Cd probes is so low, $\sim 10^{-11}$, that they do not make binary phases with other constituents. An appropriate system is a binary compound like $GdAl_2$ into which ~1 at.% of a solute dissolves onto one or other or both sublattices. Extensive unsuccessful efforts to identify this kind of ternary phase were made before $GdAl_2$(Ag) was found. This work is ongoing. Studies in $GdAl_2$ using other solutes are planned. A goal of such studies will be to elucidate rules governing interaction enthalpies between solutes on the various sublattices of a compound and to predict whether interactions are attractive or repulsive and how they depend on the nature of the host and solute elements.

GdAl$_2$ has the cubic Laves Cu$_2$Mg structure, with Gd- and Al-sites in the perfect crystal having $\overline{4}3m$ (cubic) and $\overline{3}m$ (axial) point symmetries, respectively [8]. The cubic site has zero electric field gradient (EFG) and therefore a quadrupole interaction frequency of zero. The axial site has a nonzero quadrupole interaction frequency. In the present work, samples of GdAl$_2$ were doped with [111]In activity and ~1-2 at.% of Ag. Quadrupole interactions were observed for In-probes forming complexes with neighboring solute atoms. While PAC measurements of quadrupole interactions are made for the 247 keV PAC level of the daughter [111]Cd nuclide, it should be noted that it is the interaction enthalpy between the [111]In parent probe and solute atom that is measured. Temperature dependences of the ratio of site fractions of probes with and without a solute neighbor were plotted and interaction enthalpies were determined by fitting to eq. 1. This enthalpy is the change in enthalpy when the solute atom is close to, or far away from, the probe.

## Experimental

Samples of GdAl$_2$ were doped with [111]In activity (mole fraction ~ 10$^{-11}$) and ~1-2 at.% of Ag by melting together all constituents under argon in a small arc-furnace. Metal purities were 3N for Gd and 4N for Al, and the [111]In activity was carrier-free. For each solute concentration, two samples were made that were modestly Gd-poorer or Al-poorer than the stoichiometric composition. The difference was expected to promote occupation of one sublattice or the other by the solute atoms and probes according to the heuristic rule. Sample compositions Gd$_{32.0}$Al$_{67.0}$Ag$_{1.0}$ and Gd$_{33.5}$Al$_{65.5}$Ag$_{1.0}$ were determined from masses of the elements measured prior to melting, and taking into account small mass losses during melting.

[111]In decays to the second excited state of [111]Cd by electron-capture with a mean life of 4.0 days. [111]Cd subsequently decays to the ground state with emission of 173 and 247 keV gamma rays, the 247 keV PAC level having a mean life of 120

ns.   PAC measurements were made using a four-detector PAC spectrometer employing 1.5 x 1.0 inch $BaF_2$ scintillators.   Coincidences detected between the 173 and 247 keV gamma rays give lifetimes of the PAC level and were histogrammed over typical measurement times of one day.   For each measurement, four time-coincidence spectra were accumulated simultaneously, two each at detector angles of $180^o$ and $90^o$ relative to the sample.   After fitting and subtracting accidental backgrounds, spectra were geometrically averaged and combined to yield the experimental "PAC spectrum".   The PAC spectrum was fitted with a superposition of quadrupole perturbation functions, one for each distinct site occupied by probe atoms:

$$G_2(t) = s_0 + \sum_{n=1}^{3} s_n \cos(\omega_n t) \exp\left( -\frac{\omega_n^2}{\omega_1^2} \frac{\sigma^2 t^2}{2} \right). \tag{2}$$

The perturbation function has four terms, with the three non-zero frequency components arising from hyperfine splitting of the spin $I = 5/2$ PAC level in an EFG.   They have the properties that $\omega_1 < \omega_2 < 2\omega_1$ and $\omega_3 = \omega_1 + \omega_2$.   The frequencies depend on the quadrupole interaction strength and the asymmetry parameter of the traceless EFG tensor, $\eta \equiv \left| \dfrac{V_{xx} - V_{yy}}{V_{zz}} \right|$, in which $V_{zz}$ is its principal component, so that $0 \leq \eta \leq 1$.   The fundamental observed frequency, $\omega_1$, is a function of the magnitude of the quadrupole interaction and asymmetry parameter.   For the special case of axial symmetry, $\eta = 0$, $\omega_1 : \omega_2 : \omega_3 = 1 : 2 : 3$, and $\omega_1 = \left| \dfrac{3\pi}{10} \dfrac{eQV_{zz}}{h} \right|$, in which $Q$ is the quadrupole moment of the PAC level and $h$ is Planck's constant.   For each signal, the amplitudes sum to unity, $\sum_{n=0}^{3} s_n = 1$.

The amplitude of each fitted perturbation function (eq. 2) is the site fraction for that signal.   $\sigma$ describes inhomogeneous broadening of the signals due to weak EFG disturbances from distant defects.   $\omega_1$ is used below to label the observed signals.   For additional information about PAC spectroscopy and methodology, see [9].

# Results

Six signals were detected in Ag-doped GdAl$_2$ and found consistently in many measurements. They are listed in Table 1 and identified below by average fundamental quadrupole interaction frequencies $\omega_1$. The 44, 0 and 9 Mrad/s signals were observed in the earlier study of site preferences in undoped GdAl$_2$ [1], where they were identified instead by their frequencies measured at room temperature, 49, 0 and 8 Mrad/s, respectively. The dominant 0 and 44 Mrad/s signals were attributed to isolated In-probes on Gd- and Al-sites, consistent with the point symmetries of the sites [1]. The 9 Mrad/s signal was only observed in Gd-poor samples and was reasonably attributed to an In-probe on Gd-site with a neighboring $Al_{Gd}$ antisite defect. The 54, 63 and 76 Mrad/s signals were not observed in the previous study and are attributed to close neighbor complexes of In-probes with Ag-solutes. The table gives attributions of the signals to complexes on the basis of composition dependences presented in Fig. 2 below. Also listed in the table are the types of samples in which the signals were observed and measured interaction enthalpies.

Table 1. Quadrupole interaction parameters $\omega_1$ (average value over all measurements) and $\eta$. Column 3 lists samples in which the signal is observed. Column 4 gives attributions discussed in the text. Column 5 gives the interaction enthalpy $Q$ between the two impurity atoms based on fits to eq. 1 (with a negative sign indicating an attractive interaction) and have uncertainties of order 0.04 eV.

| $\omega_1$ (Mrad/s) | $\eta$ | Samples | Attribution | Interaction Enthalpy |
|---|---|---|---|---|
| 44 | ~0 | Pure and doped | $In_{Al}$ | - |
| 0 | ~0 | Pure and doped | $In_{Gd}$ | - |
| 9 | ~0 | Pure and doped, Gd-poor | $In_{Gd} + Al_{Gd}$ | - 0.12 eV (pure); +0.15 eV (doped) |
| 54 | ~0 | Ag-doped, Gd-poor | $In_{Gd} + Ag_{Gd}$ | +0.13 eV |
| 63 | 0.3 | Ag-doped, Gd-rich | $In_{Al} + Ag_{Al}$ | - 0.10 eV |
| 76 | ~0 | Ag-doped, Gd-rich and poor | $In_{Gd} + Ag_{Al}$ | +0.17 eV |

Representative spectra are shown in Fig. 1, with the time-domain PAC spectrum on the left. (The data at apparent negative times in the figure are independent data that "mirror" the positive delayed-coincidence time spectrum. They are obtained by gating times of arrival of the 173 and 247 keV gammas in reverse order. Such "double-sided" spectra make the shape of the spectrum near time zero much clearer.) Frequency transforms are shown at right, with tridents indicating the three frequency components of the 44 and 54 Mrad/s signals. The results are now presented and discussed in several paragraphs.

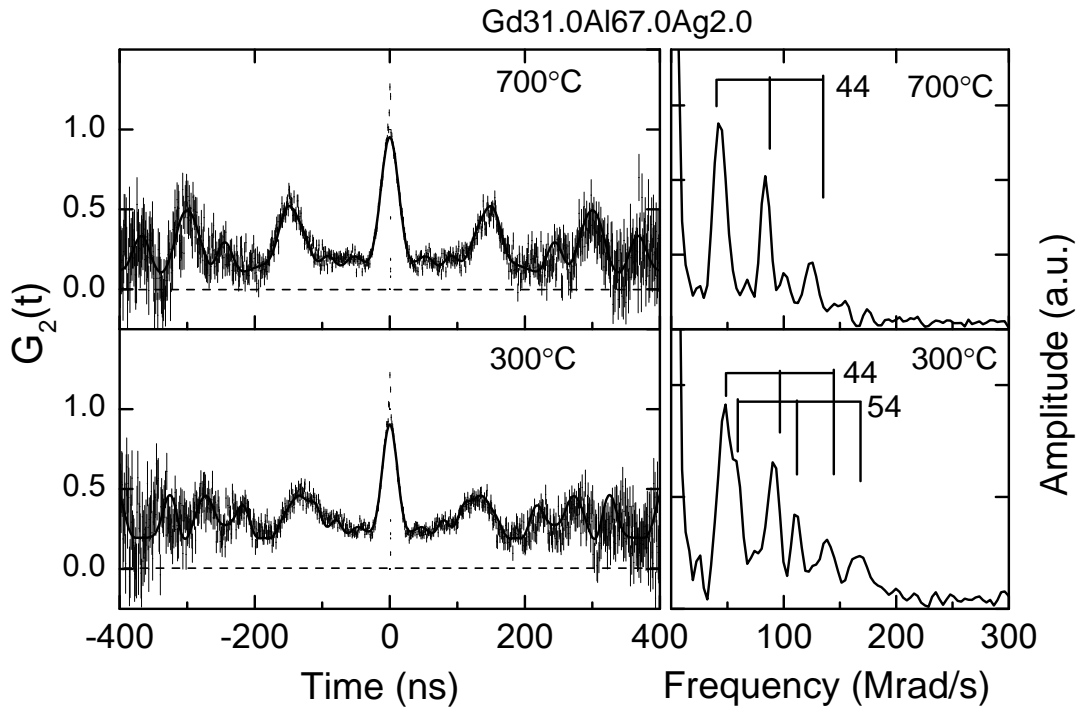

Figure 1. Representative PAC spectra for a Gd(31)Al(67)Ag(2) sample measured at 300 and 700°C, with time-domain spectra on the left and frequency spectra on the right. Two significant signals are identified at right by tridents that indicate the three frequency components. In addition, a vertical offset representing a zero-frequency signal is seen to be greater in the spectrum at 300°C.

## A.    Boundary compositions of the GdAl2 phase.

$GdAl_2$ appears on binary phase diagrams as a "line compound", having a phase field whose width has not been measured and is probably less than about 1 at.%. The more Gd-rich boundary composition is likely to be close to the 33.3% stoichiometric composition since no signal was observed in the previous work [1] that could be attributed to a $Gd_{Al}$ antisite defect. This is not unexpected since the atomic volume of Gd is much greater than of Al, leading to a large strain interaction and a high expected defect formation enthalpy for the antisite defect. The Gd-poorer boundary composition is likely to have a fixed value below the stoichiometric composition in the range 32-33.3 at.% Gd.

## B. Attributions of signals to specific probe-solute complexes based on the composition dependence of site fractions.

Fig. 2 shows fitted site fractions of the observed signals measured at 400°C plotted versus nominal average compositions of the alloys. It can be seen that some sample compositions probably lie outside the likely boundary compositions discussed above in (A). This should lead to small volume fractions of secondary phases that would produce small site fractions of additional signals that go undetected. However, no such signals were observed in this or in the previous study [1]. There follow rationales for attributions of the 54, 9, 63 and 76 Mrad/s signals.

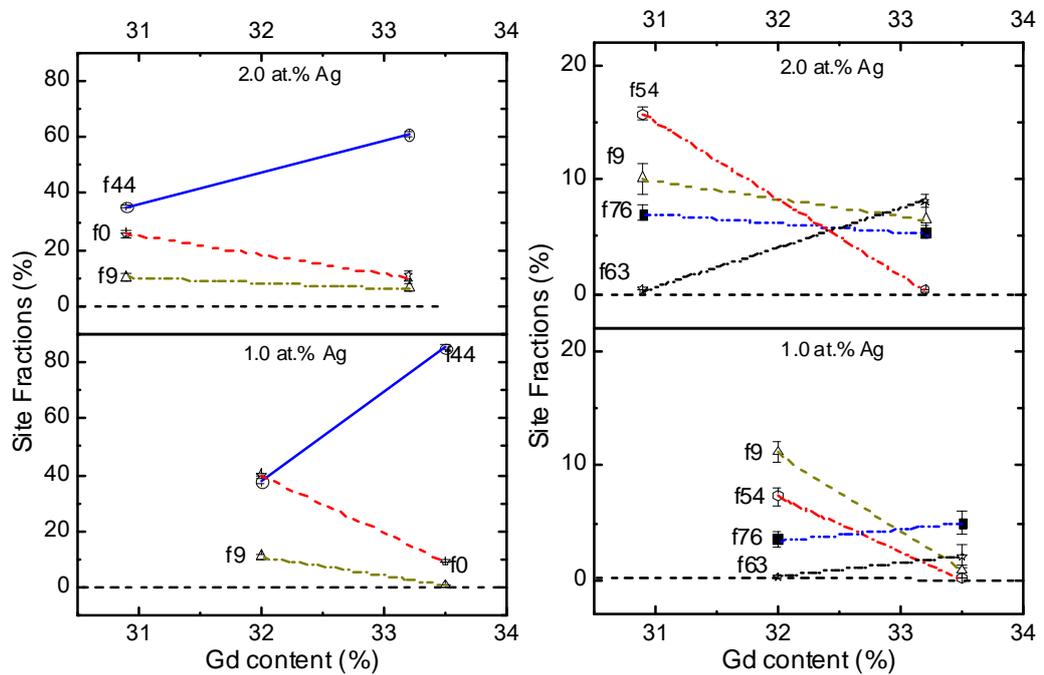

Figure 2. Site fractions of signals observed in Gd-richer and Gd-poor samples having either 1.0 or 2.0 at.% of Ag-solute. Signals are identified by quadrupole interaction frequencies given in Table 1; thus, "f9" is the site fraction for the 9 Mrad/s signal. The figure at left shows the composition dependence of the 44, 0 and 9 Mrad/s signals. At right are shown dependences for the three

signals associated with Ag-solutes as well as for the 9 Mrad/s signal previously attributed [1] to the complex $In_{Gd} + Al_{Gd}$ .

### *The 54 and 9 Mrad/s defects with $\eta = 0$ are attributed to defects on the Gd-sublattice neighboring $In_{Gd}$ probes.*

The 54 Mrad/s defect was observed only in the Gd-poorer samples.  Similarly, the site fraction of the 9 Mrad/s signal is significantly smaller in Gd-richer samples.  Both signals have axial symmetry (zero asymmetry parameter), consistent with a single defect near an $In_{Gd}$ probe.  The 9 Mrad/s signal was previously observed in undoped GdAl2 and identified with a $In_{Gd} + Al_{Gd}$ complex [1].  Appearing only in Ag-doped samples, the 54 Mrad/s signal is similarly identified with a $In_{Gd} + Ag_{Gd}$ complex.  Magnitudes of the frequencies (or EFGs) are also consistent with the small effective charge of an $Al_{Gd}$ defect (Al and Gd have the same nominal valences) and a larger anticipated effective charge for an $Ag_{Gd}$ defect (Ag and Gd have valences differing by 2).

### *The 63 Mrad/s, $\eta = 0.3$ defect is attributed to a close neighbor $In_{Al} + Ag_{Al}$ pair.*

The 63 Mrad/s site fraction disappears at the Gd-poorer phase boundary in both Ag-doped samples.   Site fractions for both the 44 and 63 Mrad/s become smaller at the Gd-poorer boundary composition, which is consistent with this attribution.  The nonaxial EFG of the complex indicates that it is a superposition of two non-collinear EFGs.  The lattice EFG at the undecorated $In_{Al}$ site is along the <111> direction in the cubic unit cell while the polar angle from an Al site to a neighboring Al-site is 25.7°, leading to a nonaxial EFG. It can be shown that all near-neighbor pairs $In_{Al} + Ag_{Al}$ will have the same EFG.

*The 76 Mrad/s signal with $\eta = 0$ is attributed to a close neighbor $In_{Gd} + Ag_{Al}$ pair.*

The site fraction of the 76 Mrad/s complex remains roughly the same at the Gd-richer and Gd-poorer boundary compositions. Having $\eta = 0$ suggests that the complex involves $In_{Gd}$ probes. At the same time, the site fraction for undecorated $In_{Gd}$ is smaller by a factor of three at the Gd-richer boundary. Only the complex $In_{Gd} + Ag_{Al}$ explains these observations satisfactorily: although the concentration of $In_{Gd}$ decreases as the composition becomes more Gd-rich, the concentration of $Ag_{Al}$ increases.

A fourth anticipated but unobserved complex is $In_{Al} + Ag_{Gd}$. It would be expected to have site fractions that were similar in Gd-rich and Gd-poor samples, in the same way as for the $In_{Gd} + Ag_{Al}$ complex with 76 Mrad/s frequency, but no such signal was identified.

## C.    Competition of host and solute atoms to occupy sites.

In pure GdAl$_2$, measurements were made on five samples of slightly different composition. An average activation enthalpy +0.343(3) eV was observed for the transfer of In-probes from Gd-sites to Al-sites [1]. Labeling the sites with their quadrupole interaction frequencies 0 and 44 Mrad/s, respectively, the temperature dependence of the site-fraction ratio is given by

$$f_0 / f_{44} = f_{Gd} / f_{Al} \sim \exp(Q / k_B T), \qquad (3)$$

in which $Q$ is an enthalpy of transfer. Figure 3 shows corresponding Arrhenius plots of the ratio of site-fractions of In-probes on Gd- and Al-sites for four samples having 1 or 2 at.% of Ag-solute and that were Gd-poor (~31.3 at.%Gd) or Gd-rich (~33.3 at.%Gd).

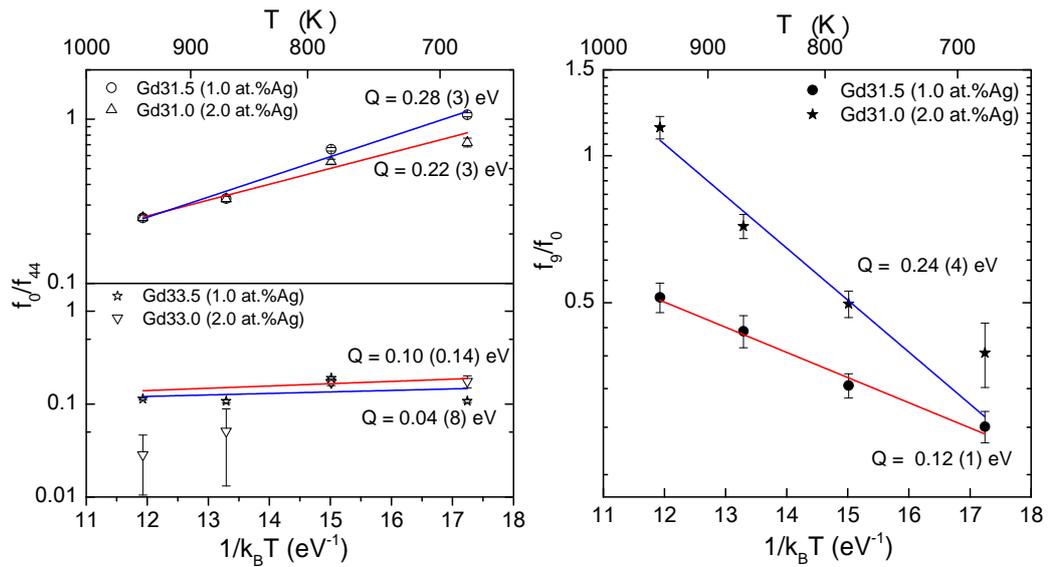

Figure 3. (Left) Arrhenius plots of the ratio of site fractions of In-probes on Gd and Al sites for samples with 1 or 2 at.% Ag and which were more Gd-poor (top) or Gd-rich (bottom). The transfer enthalpies are of order 0.25 eV (top) and 0.07 eV (bottom), much smaller than the value 0.343 eV observed for undoped GdAl$_2$ [1]. (Right) Arrhenius plots of the ratio of site fractions of the $In_{Gd} + Al_{Gd}$ complex and undecorated $In_{Gd}$ probe for samples with 1 and 2 at.% Ag.

It can be seen in Fig. 3 (left) that the corresponding activation enthalpy is ~0.25 eV for Ag-doped, Gd-poor samples and ~0.07 eV for Ag-doped, Gd-rich samples. The measured transfer enthalpies are compared in Figure 4.

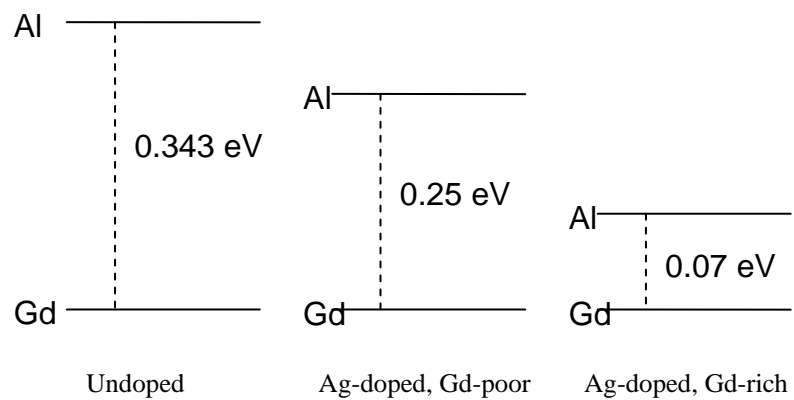

Figure 4. Schematic of measured transfer enthalpies in three types of GdAl$_2$ samples.

Naïvely, the transfer enthalpy should be equal to the difference in enthalpies of the solute at the two sites. But site-enthalpies of isolated solutes like these In-atoms depend solely on local atomic environments within 1-2 atomic shells, and therefore should be essentially independent of the mole fractions of solutes or of other defects such as antisite atoms or lattice vacancies. The large observed changes demonstrate that the transfer enthalpy is not simply equal to the difference in site-enthalpies. To explain the changes in the transfer enthalpies shown in Fig. 4, it is proposed that degeneracies of the Gd and Al-levels are strongly temperature-dependent, so that the activation enthalpies for transfer of In-solutes between the two sublattices shown in Fig. 3 or 4 are not simply equal to the difference in site-enthalpies. Eq. 3 can be is generalized to include degeneracies of sites available to In-solutes on the two sublattices:

$$\frac{f_0}{f_{44}} = \frac{f_{Gd}}{f_{Al}} \cong \frac{g_{Gd}(T)}{g_{Al}(T)} \exp(Q/k_B T) \approx \exp(Q'/k_B T) \,, \qquad (4)$$

in which the $g(T)$'s are temperature-dependent degeneracies of the two sets of levels. The observed reductions in the transfer enthalpy indicate that the ratios $g_{Gd}(T) / g_{Al}(T)$ of the degeneracies must *increase* with temperature. If one assumes that $g_{Gd}(T)$ is constant, then $g_{Al}(T)$ must *decrease* with increasing temperature in the Ag-doped samples. This is illustrated in Table 2, which lists the measured transfer enthalpies, reductions in transfer enthalpies below the value in undoped GdAl$_2$ (which is possibly equal or close to the difference in enthalpies of In on the two sites), and ratios of degeneracy factors calculated for 700K and 950K, the approximate minimum and maximum temperatures of measurement.

Table 2. Transfer enthalpies Q from Fig. 3, summarized in Fig. 4, reductions in transfer enthalpies below the value for undoped GdAl2, and ratios of degeneracy factors for Al- and Gd-sites at 950K and 700K.

| Sample | Q (eV) | ΔQ (eV) | $\dfrac{g_{Al}(950K) / g_{Al}(700K)}{g_{Gd}(950K) / g_{Gd}(700K)}$ |
|---|---|---|---|
| Undoped [ref. 1] | 0.343 | (0) | (1) |
| Ag-doped, Gd-poor | 0.25 | 0.09 | 0.67 |
| Ag-doped, Gd-rich | 0.07 | 0.27 | 0.31 |

The table shows that, relative to the undoped samples, the ratio of degeneracies has changed by factors of 1.5 and 3.2 between 750 and 900 K in the two Ag-doped samples. This is consistent with the change in compositions of the samples; the effect of increasing the average Gd-composition is to create $Gd_{Al}$ antisite atoms, thereby "filling in" virtual vacancies on the Al-sublattice and impeding transfer of Ag-atoms to the Al-sublattice. The same effect would occur in the undoped sample, but the $10^{-11}$ mole fraction of In-solutes is so minuscule that their transfer is not impeded. For Ag-doped samples, Ag-solutes switch from Gd-sites to Al-sites with increasing temperature, so that fewer Al-sites are available for In-solutes. This implies that Ag- and In-solutes have the same ordering of site-energies. If the ordering were opposite, with Ag-solutes transferring from Al-sites at low temperature to Gd-sites at high temperature, then the effective transfer enthalpy of In-solutes would have increased relative to undoped GdAl2. The fact that the reduced transfer enthalpies in the Ag-doped samples were approximately the same for 1 and 2% mole fractions of Ag indicates that 1 at.% Ag is sufficient to cause the reduction in the degeneracy ratio.

Fig. 3 (right) shows an Arrhenius plot of the ratio of site fractions of the $In_{Gd} + Al_{Gd}$ complex and undecorated $In_{Gd}$ probe for Gd-poor samples having 1 and 2 at.% Ag and ~31.3 at.% Gd. The interaction enthalpies of +0.24 eV (2% Ag) and +0.12 eV (1% Ag) can be compared with -0.16(4) eV (0% Ag) observed for an undoped GdAl2 sample with composition 33.3(1) at.% Gd [sample B in ref. 1]. The change in sign and increase in transfer enthalpy in going from 0 to 1 to 2 at.% Ag is considered to result from cross-interaction of Ag-solutes and host Al-atoms. It was already shown above that Ag atoms transfer from Gd-sites to Al-sites with increasing temperature. This transfer in turn may displace Al-atoms from their sublattice to the Gd-sublattice. If the increase in concentration of $Al_{Gd}$ with temperature is sufficiently large, it may change the sign of the effective interaction enthalpy of the complex $In_{Gd} + Al_{Gd}$ from -0.12 eV (attractive) to +0.12 to 0.24 eV (an apparently repulsive interaction), solely due to an increase in the mole fraction of $Al_{Gd}$ antisite defects. The difference in the interaction enthalpies gives the effective activation enthalpy for the increase in the concentration of $Al_{Gd}$, or about +0.24 eV (1 at.% Ag) to +0.36 eV (2% Ag). This scenario appears to explain the observations. As an alternative explanation, the ~9 Mrad/s signal might be composite and include signals having similar low frequencies from both $In_{Gd} + Al_{Gd}$ and the $In_{Al} + Ag_{Gd}$ complex not yet accounted for. But it is unlikely that the frequency would be so much lower than the 44 Mrad/s frequency of an isolated $In_{Al}$ probe atom.

### D.    Probe-solute interaction enthalpies.

Interaction enthalpies for the three In-Ag pairs were obtained by fitting ratios of site-fractions according to eq. 1. Consider first the interaction between members of the $In_{Gd} + Ag_{Gd}$ pair of solute atoms, with 54 Mrad/s frequency. The activation enthalpy to form the pair can be determined from the temperature dependence of the ratio $f(In_{Gd} + Ag_{Gd})/f(In_{Gd})$, or $f_{54}/f_0$, using Eq. 1. This is shown in Figure 5 for Gd-poor samples with 1 or 2 at.% Ag. The average activation enthalpy was +0.13 eV, indicating a repulsive interaction. One notes that the site fraction is approximately twice larger for $c$= 2% of solute than 1%, as expected from Eq. 1.

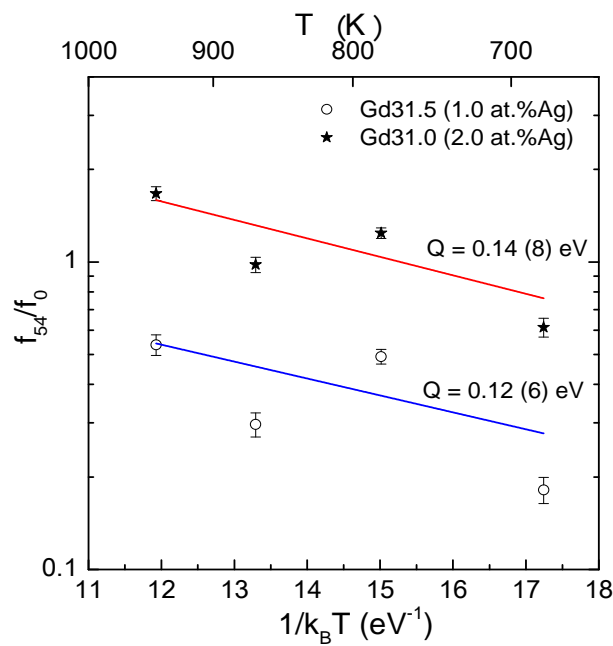

Figure 5. Arrhenius plot of the ratio of site fractions of complex $In_{Gd} + Ag_{Gd}$ and isolated probe $In_{Gd}$. Interaction enthalpies Q were obtained by fitting with Eq. 1 and are positive, indicating a repulsive interaction.

Consider as a second example the Arrhenius plot of the ratio of site fractions for the complex $In_{Gd} + Ag_{Al}$ (76 Mrad/s) and isolated probe $In_{Gd}$ (0 Mrad/s) shown in Fig. 6. The interaction enthalpies are all of order +0.17 eV, indicating a repulsive interaction.

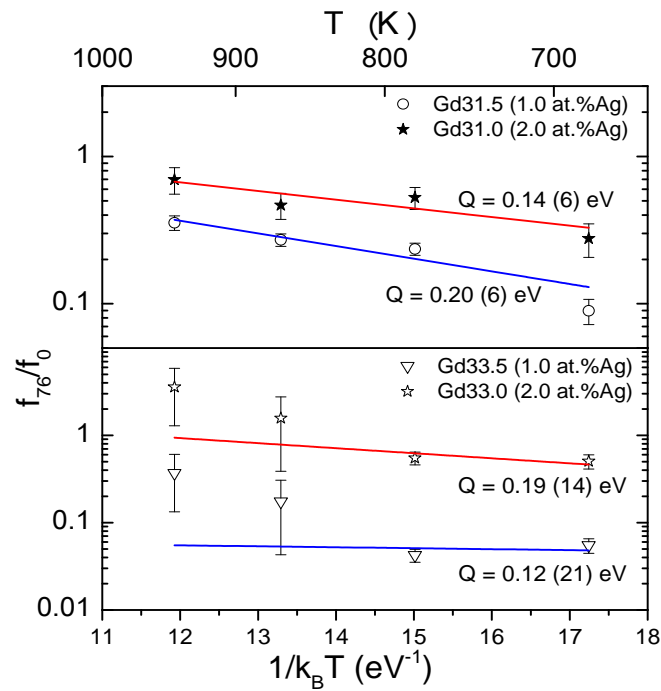

Figure 6. Arrhenius plot of the ratio of site fractions of complex $In_{Gd} + Ag_{Al}$ and isolated probe $In_{Gd}$. Interaction enthalpies Q were obtained by fitting with Eq. 1 and are positive, indicating a repulsive interaction.

Table 1 lists all the interaction enthalpies, including for the third Ag-In complex, $In_{Al} + Ag_{Al}$, which has an attractive interaction, and for the intrinsic Al-antisite complex, $In_{Gd} + Al_{Gd}$, discussed in Section C above. The magnitudes of all interaction enthalpies are small, in the range ±0.15 eV, typical of what has been observed or calculated for PAC and Mössbauer studies in pure metals [2-7].

## Summary and Conclusions.

Experiments were carried out to determine solute-solute interactions in the intermetallic compound $GdAl_2$ using PAC spectroscopy. One solute was the [111]In PAC probe, present at a mole fraction of $10^{-11}$. The other solute was 1-2 at.% of Ag. Three different close-atom pairs of Ag and In solutes were detected by characteristic EFGs. Identification of the pairs was made by examining the observed symmetry of the EFGs and how their site-fractions varied with composition. In and Ag solutes were found to populate both Gd- and Al-sublattices. Interaction enthalpies of the pairs were attractive or repulsive, and in the range $\pm 0.15$ eV, typical of what has been seen in the pure metals.

In a previous study of undoped GdAl2, a transfer enthalpy for In-solutes of +0.343 eV was observed between sites on the Gd- and Al-sublattices. In the present study, Ag-doping was found to lead to a large reduction in the effective transfer enthalpy for In-solutes between Gd- and Al-sublattices. It is proposed that this reduction is a consequence of large changes with temperature of the degeneracies of levels available for In-solutes on Gd- or Al-sublattices, with changes in the degeneracies caused by thermally-activated transfer of Ag-solutes also between Gd- and Al-sublattices. The observed trends are consistent with Ag-solutes also tending to transfer from the Gd-sublattice at low temperature to the Al-sublattice at high temperature. Through measurements such as this, it is possible to determine the direction of flow of a solute in a compound with increasing temperature.

## Acknowledgement.

This work was supported in part by the National Science Foundation under grant DMR 14-10159 (MMN Program). Krystal Kasal (MS, 2015) helped carry out early experiments to identify appropriate ternary systems for further study.

1  Matthew O. Zacate and Gary S. Collins, Physical Review B69, 174202(1-9) (2004).
2  A.Z. Hrynkiewicz and K. Krolas, Physical Review B28, 1864-1869 (1983).
3  M. Sternik and K. Krolas, Physical Review B40, 4171-4174 (1989.
4  K. Krolas, W. Bolse and L. Ziegeler, Hyperfine Interactions 35, 635 (1987).
5  M. Sternik and K. Królas, Acta Physica Polonica A82, 975-981 (1992).
6  T.E. Cranshaw, J. Phys. F: Met. Phys. 17, 1645-1657 (1987).
7  J. Chojcan, Journal of Alloys and Compounds 264, 50-53 (1998).
8  P. Villars and L.D. Calvert, Pearson's Handbook of Crystallographic Data for Intermetallic Compounds, 2$^{nd}$ ed. (ASM International, Materials Park, Ohio, 1991).
9  G. Schatz and A. Weidinger, Nuclear Condensed-Matter Physics, (John Wiley, New York, 1996).